\documentclass[letter]{IEEEtran}
\renewcommand{\baselinestretch}{0.99}

\usepackage{pgfplots}
\usepackage{hyperref}

\usetikzlibrary{shapes.multipart,intersections}
\usepackage{cite}
\usepackage{amsmath,amssymb,amsfonts,steinmetz}
\usepackage{algorithmic}
\usepackage{graphicx}
\usepackage{textcomp}
\usepackage{xcolor}
\def\BibTeX{{\rm B\kern-.05em{\sc i\kern-.025em b}\kern-.08em
    T\kern-.1667em\lower.7ex\hbox{E}\kern-.125emX}}

\usepackage{tikz}
\usetikzlibrary{calc}
\makeatletter
\newcommand{\gettikzxy}[3]{%
  \tikz@scan@one@point\pgfutil@firstofone#1\relax
  \edef#2{\the\pgf@x}%
  \edef#3{\the\pgf@y}%
}
\makeatother

\usepackage[T1]{fontenc}
\usepackage[latin9]{inputenc}
\usepackage{bm}
\usepackage{amsmath}
\usepackage{subcaption}
\usepackage{amssymb}
\usepackage{stmaryrd}
\usepackage{babel}
\usepackage{comment}
\usepackage{graphics, graphicx}
\usepackage{algorithm,algorithmic}
\usepackage{xcolor}
\usepackage{gensymb}
\usepackage{cite}
\usepackage{enumitem}
\usepackage{url}

\usepackage{appendix}
\usepackage{amsthm}
\renewcommand{\a}{\mathbf{a}}
\renewcommand{\b}{\mathbf{b}}
\renewcommand{\c}{\mathbf{c}}

\newcommand{\h}{\mathbf{h}}

\newcommand{\n}{\mathbf{n}}

\newcommand{\p}{\mathbf{p}}
\newcommand{\q}{\mathbf{q}}

\newcommand{\s}{\mathbf{s}}

\newcommand{\w}{\mathbf{w}}

\newcommand{\y}{\mathbf{y}}

\newcommand{\F}{\mathbf{F}}

\newcommand{\I}{\mathbf{I}}

\newcommand{\M}{\mathbf{M}}
\newcommand{\N}{\mathbf{N}}

\renewcommand{\P}{\mathbf{P}}
\newcommand{\Q}{\mathbf{Q}}
\newcommand{\R}{\mathbf{R}}

\newcommand{\W}{\mathbf{W}}

\newcommand{\Y}{\mathbf{Y}}

\newcommand{\Compl}{\mbox{$\mathbb{C}$}}

\renewcommand{\Re}{\mathrm{Re}}

\DeclareMathAlphabet\mathbfcal{OMS}{cmsy}{b}{n}

\newtheorem{proposition}{\bf Proposition}

\begin{document}
\title{Near-Field Localization with Dynamic Metasurface Antennas at THz: A CRB Minimizing Approach}
\author{Ioannis Gavras,~\IEEEmembership{Student~Member,~IEEE}, and George C. Alexandropoulos,~\IEEEmembership{Senior~Member,~IEEE}
\thanks{This work has been supported by the SNS JU project TERRAMETA under the EU's Horizon Europe research and innovation programme under Grant Agreement No 101097101, including top-up UKRI funding under the UK government's Horizon Europe funding guarantee.}
\thanks{The authors are with the Department of Informatics and Telecommunications,
National and Kapodistrian University of Athens, 16122 Athens, Greece. G. C. Alexandropoulos is also with the Department of Electrical and Computer Engineering, University of Illinois Chicago, IL 60601, USA (e-mails: \{giannisgav, alexandg\}@di.uoa.gr).}
}

\maketitle
\begin{abstract}
The recent trend for extremely massive antenna arrays and high frequencies facilitates localization and sensing, offering increased angular and range resolution. In this letter, we focus on the emerging technology of Dynamic Metasurface Antennas (DMAs) and present a novel framework for the design of their analog beamforming weights, targeting high accuracy near-field localization at the THz frequency band. We derive the Cram\'{e}r-Rao Bound (CRB) for the estimation {of the positions of multiple users} with a DMA-based receiver, which is then utilized as the optimization objective for the receiver's discrete tunable states of its metamaterials. Leveraging the DMA structure, we reformulate the localization objective into a constrained Rayleigh quotient maximization problem, which is efficiently solved {via two schemes: one based on projection and a greedy one.} Our simulation results verify the validity of our near-field localization analysis, showcasing the effectiveness of the proposed near-field localization designs over the optimum exhaustive search solution and state-of-the-art schemes.  
\end{abstract}

\begin{IEEEkeywords}
Dynamic metasurfaces antennas, beamforming, Cram\'{e}r-Rao bound, localization, near-field, THz.
\end{IEEEkeywords}

\vspace{-0.1cm}
\section{Introduction}
Extremely large antenna arrays are expected to be widely deployed in next generation wireless networks~\cite{XLMIMO_tutorial,41,hua2024near}, especially for high-frequency systems~\cite{THs_loc_survey}, offering numerous spatial degrees of freedom for communications, localization, and sensing applications. Dynamic Metasurface Antennas (DMAs) constitute a recent power- and cost-efficient hybrid analog and digital transceiver architecture utilizing arbitrary large numbers of phase-tunable metamaterials, which are grouped in disjoint microstrips each attached to a Radio Frequency (RF) chain and are capable for realizing analog transmit or Receive (RX) BeamForming (BF)~\cite{Shlezinger2021Dynamic}.  

The beam focusing capability of a DMA-based transmitter serving multiple User Equipment (UE) lying within its near-field regime was optimized in~\cite{zhang2022beam}. An autoregressive attention neural network for non-line-of-sight user tracking with an RX DMA was devised in~\cite{Nlos_DMA}, while \cite{NF_beam_tracking} presented a
near-field beam tracking framework for the same RX architecture. An approach for configuring the analog DMA weights for near-field localization was proposed in~\cite{yang2023near}, by reformulating the original localization objective as a received signal strength improvement one. Very recently, in~\cite{FD_HMIMO_2023,spawc2024}, focusing on near-field scenarios in the THz frequency band, designs for the digital and analog BF matrices of DMA-based full duplex transceivers were presented targeting integrated sensing and communications. However, unlike other BF architectures, such as networks of phase shifters and movable antennas~\cite{wang2023doa,lin2021hybrid,qin2024cramer}, and to the best of our knowledge, the explicit optimization of DMAs for near-field localization has not yet been studied.
%Notably, DMAs have pioneered advancements in sub-THz communications \cite{FD_HMIMO_2023,spawc2024}, capitalizing on the benefits of short wavelengths to achieve exceptional spatial resolution in localization systems. This emphasis on THz communications places us predominantly within the near-field regime \cite{THs_loc_survey}, where intricate control over the spatial characteristics of transmitted and received signals becomes paramount, necessitating highly directional and focused beams to ensure accuracy in both communication and sensing operations. However, in the context of DMA-facilitated receivers (RX), the localization performance remains uninvestigated, unlike other BeamForming (BF) architectures, such as networks of phase shifters and movable antennas.

In this letter, we study the optimization of DMA-based reception for {multi-UE} near-field localization, focusing on THz frequencies. We first derive the Cram\'{e}r-Rao Bound (CRB) for the estimation of the {UEs'} range and angular parameters as well as the position error bound (PEB), and then utilize the minimization of the former as the objective for the design of the DMA analog RX BF weights. Leveraging the partially-connected analog BF architecture of the RX DMA, we reformulate its design objective into a Rayleigh quotient optimization problem with discrete constraints onto the analog BF weights, which is efficiently solved. Our simulation results on a sub-THz narrowband channel verify the validity of our near-field localization analysis, showcasing the superiority of our presented localization-optimized DMA design over conventional ones, when all combined with Maximum Likelihood Estimation (MLE).

%This letter presents a DMA-facilitated Single-Input Multiple-Output (SIMO) system, operating in the sub-THz frequency band. Emphasizing on the near-field setting, where a single-antenna User Equipment (UE) engages in Uplink (UL) communication, 
%we derive the Cram\'{e}r-Rao Bound (CRB) for the UE's spatial parameters and design the analog RX combiner with the objective of minimizing the CRB. %Considering the tunable frequency response of the DMA RX panel, we propose both an optimal and sub-optimal solution to optimize the analog BF.  Leveraging the partially-connected structure of DMAs, allows us to deduce the CRB minimization problem into a Rayleigh quotient optimization problem that has a known optimal solution, while we also propose a sub-optimal, yet computationally efficient approach based on a greedy criterion. Sub-THz numerical simulations validate the effectiveness of these solutions, demonstrating superior Position Error Bound (PEB) performance compared to established frameworks, with comparable results between the optimal and sub-optimal solutions. Furthermore, by leveraging the Maximum Likelihood Estimator (MLE), we showcase our proficiency to attain PEB performance. Lastly, in the context of DMAs an extreme instance occurs when each microstrip comprises only one metamaterial, limiting us to a Fully Digital Antenna (FDA) configuration. Notably, our approach performs well against FDA arrangements, even with a limited number of metamaterials.

\textit{Notations:}
Vectors and matrices are represented by boldface lowercase and uppercase letters, respectively. The transpose, Hermitian transpose, inverse, and Euclidean norm are denoted as $(\cdot)^{\rm T}$, $(\cdot)^{\rm H}$, $(\cdot)^{-1}$, and $\|\cdot\|$, respectively. $[\mathbf{A}]_{i,j}$ and $[\mathbf{A}]_{i:j,u:v}$ give respectively $\mathbf{A}$'s $(i,j)$th element and sub-matrix spanning rows $i$ to $j$ and columns $u$ to $v$. $\mathbf{I}_{n}$ and $\mathbf{0}_{n}$ ($n\geq2$) are the $n\times n$ identity and zeros' matrices, respectively, and $\boldsymbol{1}_{N}$ is an $N \times 1$ column vector of ones. 
%$[\mathbf{A}]_{i,j}$ is the $(i,j)$th element of $\mathbf{A}$,  $[\mathbf{A}]_{i:j,u:v}$ is the sub-matrix of $\mathbf{A}$ spanning rows $i$ to $j$ and columns $u$ to $v$.
$\mathbb{C}$ is the complex number set, $|a|$ is the amplitude of scalar $a$, and $\jmath\triangleq\sqrt{-1}$. $\mathbb{E}\{\cdot\}$, ${\rm Tr}\{\cdot\}$, and $\Re\{\cdot\}$ give the expectation, trace, and real part, respectively. 
%is the imaginary unit. 
%$\mathbb{E}\{\cdot\}$ is the expectation operator, ${\rm Tr}\{\cdot\}$ is the trace operation, $\Re\{\cdot\}$ is the real operator, 
$\nabla$ is the vector differential operator, $\circ$ is the Hadamard product, and $\mathbf{x}\sim\mathcal{CN}(\mathbf{a},\mathbf{A})$ indicates a complex Gaussian random vector with mean $\mathbf{a}$ and covariance matrix $\mathbf{A}$.

\vspace{-0.1cm}
\section{System and Channel Models}
%We investigate a SIMO system, where a RX node, compromised by a DMA panel~\cite{Shlezinger2021Dynamic}, engages in UL communication with a single-antenna UE positioned within its effective coverage area.
We consider an RX equipped with a DMA panel~\cite{Shlezinger2021Dynamic} wishing to localize {$U$ single-antenna} UEs placed within its effective coverage area. DMAs efficiently enable the integration of numerous sub-wavelength-spaced metamaterials, which are usually grouped in microstrips each attached to a reception RF chain (comprising a low noise amplifier, a mixer downconverting the received signal from RF to baseband, and an analog-to-digital converter), within possibly extremely large apertures~\cite{41}. By dynamically tuning the responses of metamaterials in the impinging/received signal, goal-oriented analog RX BF (in this paper, localization) can be realized.
%utilizing single-RF-fed microstrips, they achieve analog BF through dynamically tunable frequency responses in the received signals. The primary objective is the accurate localization of the UE.
\begin{figure}[!t]
	\begin{center}
	\includegraphics[width=0.45\columnwidth]{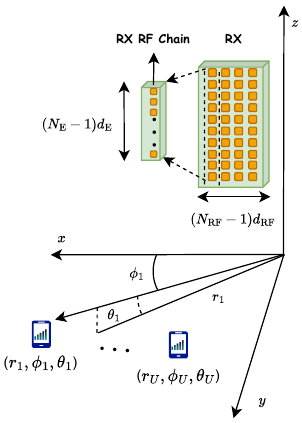}
	\caption{\small{The considered DMA-equipped RX, including its geometry and the adopted coordinate system, used for near-field localization.}}
	\label{fig: RX_DMA}
	\end{center}
\end{figure}

As depicted in Fig.~\ref{fig: RX_DMA}, the RX DMA panel is situated in the $xz$-plane with the first microstrip positioned at the origin. There exist in total $N_{\rm RF}$ microstrips within the panel, each composed of $N_{\rm E}$ distinct metamaterials placed in a uniform linear pattern with $d_{\rm E}$ distance between adjacent elements. The microstrips are individually linked to reception RF chains, which are separated from one another by a distance of $d_{\rm RF}$. Consequently, the RX DMA includes in total $N\triangleq N_{\rm RF}N_{\rm E}$ metamaterials. {The position of each $u$th ($u=1,\ldots,U$) UE is expressed in spherical coordinates as $(r_u, \theta_u, \phi_u)$, including respectively its distance from the origin of the coordinate system in Fig.~\ref{fig: RX_DMA} as well as the elevation and azimuth angles.}

We define the $N\times N$ diagonal matrix $\P_{\rm RX}$, with elements modeling signal propagation inside the RX DMA's microstrips, as follows  $\forall$$i=1,\dots,N_{\rm RF}$ and $\forall$$n = 1,\dots,N_{\rm E}$~\cite{FD_HMIMO_2023}:
\begin{align}\label{eq: TX_Sig_Prop}
    [\P_{\rm RX}]_{(i-1)N_{\rm E}+n,(i-1)N_{\rm E}+n} \triangleq \exp{(-\rho_{i,n}(\alpha_i + \jmath\beta_i))},
\end{align}
where $\alpha_i$ represents the waveguide attenuation coefficient, $\beta_i$ stands for the wavenumber, and $\rho_{i,n}$ signifies the position of the $n$th element within the $i$th microstrip. Let also $w^{\rm RX}_{i,n}$ be the adjustable response (i.e., analog weight), associated with each $n$th metamaterial in each $i$th microstrip. This weight is assumed to conform to a Lorentzian-constrained phase model and is assumed to belong to the phase profile codebook $\mathcal{W}$:
\begin{align}\label{eq: code}
    w^{\rm RX}_{i,n} \in \mathcal{W}\triangleq \{0.5\left(\jmath+e^{\jmath\phi}\right)|\phi\in\left[-\pi/2,\pi/2\right]\}.
\end{align}
Using this definition, the DMA's analog RX BF matrix $\W_{\rm RX}\in\mathbb{C}^{N\times N_{\rm RF}}$ is obtained as $[\W_{\rm RX}]_{(i-1)N_{\rm E}+n,j}=w^{\rm RX}_{i,n}$ for $i=j$, and as $[\W_{\rm RX}]_{(i-1)N_{\rm E}+n,j}=0$ for $i\neq j$.
%
%follows:
%\begin{align}
%    [\W_{\rm RX}]_{(i-1)N_{\rm E}+n,j} = \begin{cases}
%    w^{\rm RX}_{i,n},&  i=j,\\
%    0,              & i\neq j.
%\end{cases}
%\end{align} 
\vspace{-0.1cm}
\subsection{Near-Field Channel Model}
We study wireless operations in the THz frequency band and within a near-field signal propagation environment~\cite{THs_loc_survey}. {The $N$-element complex-valued vector channel between the RX DMA and each $u$th single-antenna UE is modeled as follows:
\begin{align}
    \label{eqn:UL_chan}
    [\h_u]_{(i-1)N_{\rm E}+n} \triangleq \alpha_{u,i,n} \exp\left(\frac{\jmath2\pi}{\lambda} r_{u,i,n}\right),
\end{align}
where $r_{u,i,n}$ denotes the distance from the $u$th UE to the $n$th metamaterial of the $i$th microstrip (and respective RF chain) of the RX DMA panel. The attenuation factor $\alpha_{u,i,n}$} accommodates molecular absorption with a coefficient of $\kappa_{\rm abs}$ in the THz range, and is formally expressed via the formula:
\begin{align}\label{eq: atn}
    \alpha_{u,i,n} \triangleq \sqrt{F(\theta_{u,i,n})} \frac{\lambda}{4\pi r_{u,i,n}} \exp\left(-\frac{\kappa_{\rm abs}r_{u,i,n}}{2}\right)
\end{align}
with $\lambda$ being the wavelength and $F(\cdot)$ is each metamaterial's radiation profile, modeled for an elevation angle $\theta$ as follows:
\begin{align}
    F (\theta) = \begin{cases}
    2(b+1)\cos^{b}(\theta),& {\rm if}\, \theta\in[-\frac{\pi}{2},\frac{\pi}{2}],\\
    0,              & {\rm otherwise}.
\end{cases}
\end{align}
In the latter expression, $b$ determines the boresight antenna gain which depends on the specific DMA technology~\cite{Shlezinger2021Dynamic}. {It is finally noted that each distance $r_{u,i,n}$ in \eqref{eqn:UL_chan} and \eqref{eq: atn} can be calculated, using the coordinate system in Fig.~\ref{fig: RX_DMA}, as follows:
\begin{align}\label{eq: dist}
     r_{u,i,n} = \Big(&\!(r_u\sin\theta_u\cos\phi_u -(i\!-\!1)d_{\rm RF})^2 +\\ &(r_u\sin\theta_u\sin\phi_u)^2 + (r_u\cos\theta_u\!-\!(n\!-\!1)d_{\rm E})^2\Big)^{\frac{1}{2}}.\nonumber
\end{align}
The elevation angle of each $u$th UE's antenna with respect to each $n$th element of each $i$th microstrip is expressed as:}
\begin{align}\label{eq:thetas}
    {\theta_{u,i,n} \triangleq \sin^{-1}\left(r_{u,i,n}^{-1}|(n-1)d_{\rm E}-r_u\cos{\theta_u}|\right).}
\end{align}

\subsection{Received Signal Model}
The baseband received signals at the outputs of the RX DMA's RF chains after $T$ UE pilot transmissions can be mathematically expressed via the matrix $\Y\triangleq [\y(1),\ldots,\y(T)]\in\Compl^{N_{\rm RF}\times T}$ with $t=1,\ldots,T$:
%Assuming we employ $T$ transmissions for each communication slot, where $t\in\{1,\ldots, T\}$, the mathematical expression for the baseband received signal $\Y\in\Compl^{N_{\rm RF}\times T}$, is as follows:
\begin{align}\label{eq:UL_signal_matrix}
    {\Y = \W_{\rm RX}^{\rm H}\P_{\rm RX}^{\rm H}\sum_{u=1}^U\h_u^{\rm H}\s_u+\W_{\rm RX}^{\rm H}\P_{\rm RX}^{\rm H}\N,}
\end{align}
where $\N \triangleq [\n(1),\ldots,\n(T)]\in\Compl^{N\times T}$ with each $\n(t)\sim\mathcal{CN}(\mathbf{0},\sigma^2\mathbf{I}_{\rm N_{\rm RF}})$ being the Additive White Gaussian Noise (AWGN) vector, {and $\s_u \triangleq [s_u(1),\ldots,s_u(T)]\in\Compl^{1\times T}$. Each transmitted pilot signal $s_u(t)$ is subjected to the power constraint $\mathbb{E}\{\|s_u(t)\|^2\}\leq P_{\max}$}, where $P_{\max}$ signifies the common maximum UE transmission power in the uplink. %By setting $s(t)=\sqrt{P_{\max}}$ $\forall t$, the received signal at the RX DMA at each $t$th transmission deduces to:
% \begin{align}\label{eq:UL_signal}
%     \y(t) = \W_{\rm RX}^{\rm H}\P_{\rm RX}^{\rm H}\h^{\rm H}\sqrt{P_{\max}} +\W_{\rm RX}^{\rm H}\P_{\rm RX}^{\rm H}\n(t).
% \end{align}
\vspace{-0.1cm}
\section{DMA Design for Near-Field Localization}\label{Sec: 3}
In this section, we present our RX DMA design for near-field localization. We first derive the CRB and PEB for the intended estimation, which are used for our design objective formulation. Finally, our solution for the DMA's analog BF weights is presented along with its computational complexity.
%focus on designing the parameters for the proposed massive SIMO system tailored for near-field localization. Our approach begins by deriving the CRB regarding the UE's position. Then, we determine the RX DMA array's analog BF matrix, guided by the optimization objective of minimizing the CRB.

\subsection{PEB Analysis} %using \eqref{eq:UL_signal}
{It is evident from  \eqref{eq:UL_signal_matrix}'s inspection that, for a sufficiently large number $T$ of pilot transmissions per UE within a coherent channel block, yields $T^{-1}\mathbb{E}\{\s_u\s_u^{\rm H}\}= P_{\max}\I_{N_{\rm RF}}$ $\forall u$, indicating that the received signal at the DMA's RF chains is distributed as $\Y\sim\mathcal{CN}(\boldsymbol{\M},\R_n)$ with mean $\boldsymbol{\M} \triangleq \W_{\rm RX}^{\rm H}\P_{\rm RX}^{\rm H}\sum_{u=1}^U\h_u^{\rm H}\s_u$ and covariance $\R_n \triangleq \sigma^2\W_{\rm RX}^{\rm H}\P_{\rm RX}^{\rm H}\P_{\rm RX}\W_{\rm RX}\I_{N_{\rm RF}}$. By defining the vector $\boldsymbol{\zeta} \triangleq [r_1,\theta_1,\phi_1,\ldots,r_U,\theta_U,\phi_U]^{\rm T}$ with the parameters of the true UE positions, the $3UT\times3UT$ Fisher Information Matrix (FIM) for its estimation can be calculated as follows~\cite{spawc2024}: 
\begin{align}\label{eq:FIM}
    \mathbfcal{I} =\begin{bmatrix}
    \mathbfcal{I}_{r_1r_1} & \mathbfcal{I}_{r_1\theta_1} & \mathbfcal{I}_{r_1\phi_1} & & \mathbfcal{I}_{r_1\phi_U}\\
    \mathbfcal{I}_{\theta_1 r_1} & \mathbfcal{I}_{\theta_1\theta_1} & \mathbfcal{I}_{\theta_1\phi_1} & \ldots & \mathbfcal{I}_{\theta_1\phi_U}\\
    \mathbfcal{I}_{\phi_1 r_1} & \mathbfcal{I}_{\phi_1\theta_1} & \mathbfcal{I}_{\phi_1\phi_1} & & \mathbfcal{I}_{\phi_1\phi_U} \\
    & \vdots &  & \ddots & \\
    \mathbfcal{I}_{\phi_U r_1} & \mathbfcal{I}_{\phi_U\theta_1} & \mathbfcal{I}_{\phi_U\phi_1} & \ldots& \mathbfcal{I}_{\phi_U\phi_U}
    \end{bmatrix},
\end{align}
where each $(i,j)$th ($i,j=1,\ldots,3U$) matrix element $[\mathbfcal{I}]_{i,j}$ of size $T\times T$ is given by:
\begin{align}
    \nonumber[\mathbfcal{I}]_{i,j}\!=\!2\Re\left\{\!\frac{\partial \boldsymbol{\M}^{\rm H}}{\partial[\boldsymbol{\zeta}]_i}\R_n^{-1}\frac{\partial \boldsymbol{\M}}{\partial[\boldsymbol{\zeta}]_j}\!\right\}\!+\!\text{Tr}\left\{\!\R_n^{-1}\frac{\partial\R_n}{\partial[\boldsymbol{\zeta}]_i}\R_n^{-1}\frac{\partial\R_n}{\partial[\boldsymbol{\zeta}]_j}\!\right\}\!.
\end{align}
Since $\nabla_{\boldsymbol{\zeta}}\R_n = \mathbf{0}_{3U\times1}$ due to the fact that $\R_n$ is independent of $\boldsymbol{\zeta}$, it holds that each FIM value depends solely on $\Y$'s mean value. We next compute the following derivative $\forall i$:
\begin{align}\label{eq: deriv_mean}
    \frac{\partial\boldsymbol{\M}}{\partial[\boldsymbol{\zeta}]_i} = \W_{\rm RX}^{\rm H}\P_{\rm RX}^{\rm H}\sum_{u=1}^U\frac{\partial\h_u^{\rm H}}{\partial[\boldsymbol{\zeta}]_i}\s_u.
\end{align}
Consequently, each diagonal element of the FIM matrix in \eqref{eq:FIM} can be expressed as follows with $\F\triangleq\W_{\rm RX}\R_n^{-1}\W_{\rm RX}^{\rm H}$:
\begin{align}\label{eq: FIM_diag}
    [\mathbfcal{I}]_{i,i}=2\sum_{u=1}^U\Re\left\{\s_u^{\rm H}\frac{\partial\h_u}{\partial[\boldsymbol{\zeta}]_i}\P_{\rm RX}\F\P_{\rm RX}^{\rm H}\frac{\partial\h_u^{\rm H}}{\partial[\boldsymbol{\zeta}]_i}\s_u\right\}.
\end{align}
Putting all above together, the sum-PEB with respect to all UEs' with polar coordinates $\{(r_1,\theta_1,\phi_1),\ldots,(r_U,\theta_U,\phi_U)\}$, considering an RX DMA, is given by \cite{kay1993fundamentals}:}
\begin{align}\label{eq:PEB}
\text{PEB}_{\boldsymbol{\zeta}} \triangleq 
%\sqrt{{\rm Tr}\{\text{CRB}(r,\theta,\phi)\}}=
{\sqrt{\text{CRB}_{\rm \boldsymbol{\zeta}}}=\sqrt{{\rm Tr}\left\{\mathbfcal{I}^{-1}\right\}}.}
%\sqrt{{\rm Tr}\left\{\underbrace{\left[\mathbfcal{I}_{rr}+\mathbfcal{I}_{\theta\theta}+\mathbfcal{I}_{\phi\phi}\right]^{-1}}_{\triangleq\text{CRB}(r,\theta,\phi)}\right\}}. 
\end{align}

\subsection{DMA Design Objective}\label{Sec: opt}
{Our goal, in this paper, is to optimize the RX DMA's analog BF weights for facilitating the localization of multiple UEs lying in its near-field region. To this end, we leverage the positive semidefinite nature of the FIM and the lower bound ${\rm Tr}\{\mathbfcal{I}^{-1}\}\geq\frac{9U^2T^2}{{\rm Tr}\{\mathbfcal{I}\}}$ resulting from the following proposition.}
\begin{proposition}[{Inequality for positive semidefinite matrices}]
{For any $N\times N$ complex positive semidefinite and invertible matrix $\mathbfcal{K}$, the inequality ${\rm Tr}\{\mathbfcal{K}^{-1}\}\geq\frac{N^2}{{\rm Tr}\{\mathbfcal{K}\}}$ holds. }
\begin{proof}
{Let $\lambda_1,\ldots,\lambda_N\geq0$ be $\mathbfcal{K}$'s eigenvalues, yielding the expressions ${\rm Tr}\{\mathbfcal{K}\} = \sum_{i=1}^N \lambda_i$ and $ {\rm Tr}\{\mathbfcal{K}^{-1}\} = \sum_{i=1}^N \lambda_i^{-1}$.
%the traces of $\mathbfcal{K}$ and $\mathbfcal{K}^{-1}$ are given by :
%\[
%{\rm Tr}\{\mathbfcal{K}\} = \sum_{i=1}^N \lambda_i, \quad {\rm Tr}\{\mathbfcal{K}^{-1}\} = \sum_{i=1}^N \frac{1}{\lambda_i}.
%\]
It is deduced from the harmonic-geometric mean inequality:
\[
\frac{N}{\sum_{i=1}^N \lambda_i^{-1}} \leq \frac{\sum_{i=1}^N \lambda_i}{N}\,\,\Leftrightarrow\,\,\sum_{i=1}^N \lambda_i^{-1} \geq \frac{N^2}{\sum_{i=1}^N \lambda_i}.
\]
%Taking the reciprocal of both sides, we obtain:
%\[
%\sum_{i=1}^N \frac{1}{\lambda_i} \geq \frac{N^2}{\sum_{i=1}^N \lambda_i}.
%\]
Substituting the previous trace definitions for $\mathbfcal{K}$ and its inverse in the latter inequality, completes the proof.
%, this becomes:
%\[
%{\rm Tr}\{\mathbfcal{K}^{-1}\} \geq \frac{N^2}{{\rm Tr}\{\mathbfcal{K}\}}.
%\]
%Thus, the inequality is proven.
}
\end{proof}
\end{proposition}

We now focus on the minimization of the latter lower bound for the estimation of the UEs' positions parameter vector $\boldsymbol{\zeta}$, which can be mathematically formulated via the optimization:
%Our goal in this paper is to optimize the analog RX BF weights at the RX DMA for near-field localization. To this end, leveraging the positive semidefinite nature of the FIM matrix and the previously derived PEB expression, we focus on the minimization of the CRB for the estimation of the UE position parameter vector $\boldsymbol{\zeta}$ (i.e., the input to ${\rm Tr}\{\cdot\}$ in \eqref{eq:PEB}), which can be equivalently expressed via the optimization:
\begin{align}
        {\mathcal{OP}: \nonumber\underset{\substack{\W_{\rm RX}}}{\max} \,\, {\rm Tr}\left\{\mathbfcal{I}\right\}\,\,\,\text{\text{s}.\text{t}.}\,\,\, w^{\rm RX}_{i,n} \in \mathcal{W}\,\,\forall i,n.}
\end{align}
{We next use the vector definitions $\a_u\triangleq\P_{\rm RX}^{\rm H}\frac{\partial\h_u^{\rm H}}{\partial r_u}$, $\b_u\triangleq\P_{\rm RX}^{\rm H}\frac{\partial\h_u^{\rm H}}{\partial\theta_u}$, and $\c_u\triangleq\P_{\rm RX}^{\rm H}\frac{\partial\h_u^{\rm H}}{\partial\phi_u}$, to simplify $\mathcal{OP}$'s objective as follows: ${\rm Tr}\left\{\Re\left\{\left(\sum_{u=1}^U\a_u\a_u^{\rm H}+\b_u\b_u^{\rm H}+\c_u\c_u^{\rm H}\right)\F\right\}\right\}$.}
%\begin{align}
%        &\nonumber\underset{\substack{\W_{\rm RX}}}{\max} \quad {\rm Tr}\left\{\Re\{(\a\a^{\rm H}+\b\b^{\rm H}+\c\c^{\rm H})\F\}\right\}\\\nonumber&\quad\text{\text{s}.\text{t}.}\, w^{\rm RX}_{i,n} \in \mathcal{W}.
%\end{align}

\begin{comment}
To simplify $\mathcal{OP}$ one can readily demonstrate that $\R_n$ takes the form of a block diagonal matrix with the following distinctive composition:
\begin{align}
    \nonumber\R_n = \text{diag}\{&(\w_1^H\circ\p_1^H)(\p_1\circ\w_1), \ldots ,\\
    &(\w_{N_{\rm RF}}^H\circ\p_{N_{\rm RF}}^H)(\p_{N_{\rm RF}}\circ\w_{N_{\rm RF}})\}\frac{\sigma^2}{T}
\end{align}
where $\circ$ is the Hadamard product and
\begin{align}
    &\nonumber \p_i = [\text{diag}(\P_{\rm RX})]_{((i-1)N_{\rm E}+1:iN_{\rm E})}, \quad \forall i=1,\ldots,N_{\rm RF},\\
    &\nonumber \w_i = [\W_{\rm RX}]_{((i-1)N_{\rm E}+1:iN_{\rm E},i)}, \quad \forall i=1,\ldots,N_{\rm RF}.
\end{align}
Consequently, the expression $\W_{\rm RX}\R_n^{-1}\W_{\rm RX}^H$ yields a block diagonal matrix, where each block is structured as:

\begin{align}
    \frac{\w_i\w_i^H}{(\w_i^H\circ\p_i^H)(\p_i\circ\w_i)\frac{\sigma^2}{T}} \,\quad \forall i=1,\ldots,N_{\rm RF}.
\end{align}
\end{comment}

\vspace{-0.1cm}
\subsection{Proposed Analog RX BF for Near-Field Localization}\label{our_solution}
Profiting from the partially-connected analog RX BF architecture of the DMA~\cite{Shlezinger2021Dynamic}, $\Y$'s covariance matrix $\R_n$ is a diagonal matrix with the following distinctive composition:
\begin{align}
    \nonumber\R_n = \sigma^2\text{diag}\{&(\w_1^{\rm H}\circ\p_1^{\rm H})(\p_1\circ\w_1), \ldots ,\\
    &(\w_{N_{\rm RF}}^{\rm H}\circ\p_{N_{\rm RF}}^{\rm H})(\p_{N_{\rm RF}}\circ\w_{N_{\rm RF}})\},
\end{align}
% $\circ$ is the Hadamard product 
where $\forall$$i=1,\ldots,N_{\rm RF}$ using $\kappa_i\triangleq (i-1)N_{\rm E}+1$:
\begin{align}
    \p_i \triangleq [\text{diag}(\P_{\rm RX})]_{\kappa_i:iN_{\rm E}},% \quad \forall i=1,\ldots,N_{\rm RF},\\
    \quad \w_i \triangleq [\W_{\rm RX}]_{\kappa_i:iN_{\rm E},i}.% \quad \forall i=1,\ldots,N_{\rm RF}.
\end{align}
Consequently, it can be easily verified that matrix $\F$ appearing in $\mathcal{OP}$'s simplified objective has a block diagonal structure with each block $i$ (out of the $N_{\rm RF}$) having the following form:
\begin{align}\label{eq:block}
    [\F]_{\kappa_i:iN_{\rm E},\kappa_i:iN_{\rm E}}=
    %s\frac{\w_i\w_i^{\rm H}}{(\sigma^2\w_i^{\rm H}\circ\p_i^{\rm H})(\p_i\circ\w_i)}=
    \frac{\w_i\w_i^{\rm H}}{\sigma^2\|\w_i\|^2\|\p_i\|^2}.
\end{align}

{It can be easily shown for the matrix $\Q\triangleq\sum_{u=1}^U\a_u\a_u^{\rm H}+\b_u\b_u^{\rm H}+\c_u\c_u^{\rm H}$} appearing in $\mathcal{OP}$'s simplified objective that it is a positive semidefinite matrix. Using this property and $\F$'s block diagonal structure, as shown in~\eqref{eq:block}, $\mathcal{OP}$ can be decomposed into the following $N_{\rm RF}$ sub-problems that can be solved in parallel:
\begin{align}
    \nonumber\mathcal{OP}_i:\, &\underset{\substack{\w_i}}{\max} \,\, {\rm Tr}\left\{\Re\left\{\frac{\w_i^{\rm H}[\Q]_{\kappa_i:iN_{\rm E},\kappa_i:iN_{\rm E}}\w_i}{\|\w_i\|^2}\right\}\right\}\\&\nonumber\,\,\,\text{\text{s}.\text{t}.}\,\,\, w^{\rm RX}_{i,n} \in \mathcal{W}\,\,\forall i,n.
\end{align}
Clearly, each $\mathcal{OP}_i$ deals with the maximization of a Rayleigh quotient~\cite{zhang2013optimizing}, thus, the optimal solution for each $\w_i$, when neglecting the codebook-based constraint, is obtained from $[\Q]_{\kappa_i:iN_{\rm E},\kappa_i:iN_{\rm E}}$'s principal singular vector, denoted by $\q_i$. Specifically, for each $i$th RF chain at the RX DMA, we can construct the optimal Lorentzian-constrained analog BF vector as $\bar{\w}_i\triangleq0.5(\jmath\boldsymbol{1}_{N_{\rm E}}+\q_i)$.
%, where $\boldsymbol{1}_{\rm N_{\rm E}}$ a $N_{\rm E}\times 1$ column vector of ones. 
Then, to deal with each $\mathcal{OP}_i$'s constraint, we propose to project each $\bar{\w}_i$ onto the finite-sized codebook $\mathcal{W}$, and consequently seek via an one-dimensional search for the best $N$-element analog RX BF vector $\w$ for the DMA, that is constructed from $\mathcal{W}$, minimizing the following vector distance criterion:
\begin{align}\label{eq: project}
    \w_i = \arg\min_{\w} \sqrt{1-\left|\frac{\w^{\rm H}\bar{\w}_i}{\|\w\|\|\bar{\w}_i\|}\right|}.
\end{align}
{The complete algorithm for this projection-based solution of $\mathcal{OP}$ is summarized in Algorithm~\ref{alg:the_opt1}.}

\begin{algorithm}[!t]
{
    \caption{Projection-Based Solution for $\mathcal{OP}$}
    \label{alg:the_opt1}
    \begin{algorithmic}[1]
        \renewcommand{\algorithmicrequire}{\textbf{Input:}}
        \renewcommand{\algorithmicensure}{\textbf{Output:}}
        \REQUIRE Initialization for $\W_{\rm RX}$, $\P_{\rm RX}$, $\mathcal{W}$, and initial estimate\\$\quad\,$ for ${\boldsymbol{\zeta}}$.
        \ENSURE $\W_{\rm RX}$.
               \STATE Construct $\Q$ and solve $\mathcal{OP}_i$ $\forall i$ to obtain $\q_i$'s. Then, compute each $i$th Lorentzian-constrained vector via the expression $\bar{\w}_i=0.5(\jmath\boldsymbol{1}_{N_{\rm E}}+\q_i)$.
               \STATE Project each $\bar{\w}_i$ onto the codebook $\mathcal{W}$ via an one-dimensional search using \eqref{eq: project}.% to obtain the corresponding codebook-constrained codeword $\w_i$.
    \end{algorithmic}}
\end{algorithm}

\subsection{Computational Complexity Analysis}
The optimal solution for $\mathcal{OP}$ can be obtained by exhaustive search, which is computationally demanding with $\mathcal{O}(\binom{N_{\rm \mathcal{W}}}{N_{\rm RF}})$ of complexity, where $N_{\rm \mathcal{W}}$ denotes the total number of analog RX BF vectors for the DMA that can be constructed from $\mathcal{W}$. On the other hand, our projection-based sub-optimal solution for all $\mathcal{OP}_i$s' presented in Section~\ref{our_solution} needs smaller computational complexity of $\mathcal{O}(N_{\rm RF}(N_{\rm E}^3+N_{\rm \mathcal{W}}))$. Specifically, $\mathcal{O}(N_{\rm E}^3)$ of complexity is required for the singular value decompositions of $[\Q]_{\kappa_i:iN_{\rm E},\kappa_i:iN_{\rm E}}$'s, while all projection steps need $\mathcal{O}(N_{\rm RF}N_{\rm \mathcal{W}})$ mathematical operations.

%The convergence rate of $\mathcal{OP}$ is primarily impacted by the number of metamaterials present in the DMA. 
%In the proposed approach, the computational complexity is constrained by $\mathcal{O}(N_{\rm E}N_{\rm RF}^4+N_{\rm \mathcal{W}})$. This complexity arises from executing the SVD $N_{\rm RF}$ times for square matrices, resulting in a complexity of $\mathcal{O}(N_{\rm E}N_{\rm RF}^3)$ for the SVD operation. Additionally, the projection step contributes a complexity of $\mathcal{O}(N_{\rm \mathcal{W}})$.
%Exhaustive search is bounded by a computational complexity of $\mathcal{O}(N_{\rm \mathcal{W}}N_{\rm RF})$ as it necessitates performing $N_{\rm RF}$ separate 1D searches over the codebook, corresponding to each $\mathcal{OP}_i$.

In the results that follow, we will also investigate the following greedy-based approach for solving $\mathcal{OP}$, which requires only $\mathcal{O}(N_{\mathcal{W}})$ of complexity. Starting with $N_{\rm RF}$ randomly selected analog RX BF vectors from $\mathcal{W}$, we continue with an additional random selection of a vector to replace any of the $N_{\rm RF}$ previous ones yielding the largest value for $\mathcal{OP}$'s objective function. Then, this selected vector is removed from $\mathcal{W}$. This process is repeated until all vectors from $\mathcal{W}$ are selected. It is noted that, if $N_{\rm RF}>N_{\rm \mathcal{W}}$ holds, $N_{\rm RF}$ copies of codebook $\mathcal{W}$ will be utilized for the greedy-based approach. {The complete algorithm for the greedy-based solution of $\mathcal{OP}$ is summarized in Algorithm~\ref{alg:the_opt2}.}

\begin{algorithm}[!t]
    \caption{Greedy-Based Solution for $\mathcal{OP}$}
    {\label{alg:the_opt2}
    \begin{algorithmic}[1]
        \renewcommand{\algorithmicrequire}{\textbf{Input:}}
        \renewcommand{\algorithmicensure}{\textbf{Output:}}
        \REQUIRE Initialization for $\W_{\rm RX}$, $\P_{\rm RX}$, $\mathcal{W}$, and initial estimate \\$\quad\,\,\,$for ${\boldsymbol{\zeta}}$.
        \ENSURE $\W_{\rm RX}$.
               \STATE Initialize the set $\mathcal{S}$ with $N_{\rm RF}$ randomly selected codewords from $\mathcal{W}$, and set $\overline{\mathcal{W}}=\mathcal{W}$.
               \WHILE{$\overline{\mathcal{W}}$ is not empty}
                    \STATE Select a random codeword $\bar{\w}$ from $\overline{\mathcal{W}}$ and then remove it from this codebook.
                    \STATE Determine whether $\bar{\w}$ can replace any codeword in the set $\mathcal{S}$ to achieve a larger improvement $\mathcal{OP}$'s objective. If such a replacement is possible, update $\mathcal{S}$ accordingly.
               \ENDWHILE
    \end{algorithmic}}
\end{algorithm}

%The greedy-based proposed approach is bounded by $\mathcal{O}(N_{\mathcal{W}})$, because we perform only one 1D search over the codebook 
%The convergence rate of $\mathcal{OP}$ is influenced by the quantity of metamaterials in the DMA. 
%We propose a sub-optimal but computationally efficient greedy strategy, in which we initialize a subset with $N_{\rm RF}$ codewords randomly selected from a predefined finite size codebook $\mathcal{W}$. Then we randomly choose a codeword from $\mathcal{W}$ and assess its potential to replace a single codeword in the subset, depending on whether it results in a superior gain, as determined by the $\mathcal{OP}$'s objective function. The selected codeword is removed from the codebook regardless of the replacement outcome. This process repeats until all codewords have been evaluated, resulting in the final optimized subset. 
%Here, the procedure exhibits greediness in its selection of whether a codeword will replace another within the set. It solely evaluates the current subset of codewords, without regard for past or future actions.

\vspace{-0.1cm}
\section{Numerical Results and Discussion}
In this section, we numerically evaluate the performance of the proposed RX DMA design for near-field localization, when used in conjunction with the MLE scheme~\cite{myung2003tutorial}. We have considered the scenario of Fig.~\ref{fig: RX_DMA} at $120$ GHz central frequency with a $B=150$ KHz bandwidth, including {$U=3$ single-antenna UEs} lying in the DMA's near-field (Fresnel) region given by $\theta_u = 30^{\circ}$, $\phi_u\in[1^{\circ}, 90^{\circ}]$, and $r_u\in[1,12]$ meters, $\forall u$. In the RX DMA, its microstrips and metamaterials were spaced with $d_{\rm RF}=\frac{\lambda}{2}$ and $d_{\rm E}=\frac{\lambda}{5}$, respectively. A $3$-bit beam codebook $\mathcal{W}$ was used, initially restricting all elements' phase responses to the set $\mathcal{F}\in\{e^{j\phi}|\phi\in\left[-\pi/2,\pi/2\right]\}$, having elements of constant unit amplitude and uniformly distributed phase values. Then, to compensate for the signal propagation inside the DMA microstrips, we set $w^{\rm RX}_{i,n} = 0.5(\jmath+\widetilde{w}^{\rm RX}_{i,n}e^{\jmath\rho_{i,n} \beta_{i}})$ with $\widetilde{w}^{\rm RX}_{i,n}\in\mathcal{F}$ $\forall$$i,n$. We have conducted $300$ Monte Carlo simulations, each including $T=100$ UE pilot transmissions, and set AWGN's variance as $\sigma^2=-174 + 10\log_{10}(B)$ in dBm.

\begin{figure}[!t]
\centering
\includegraphics[width=0.78\columnwidth]{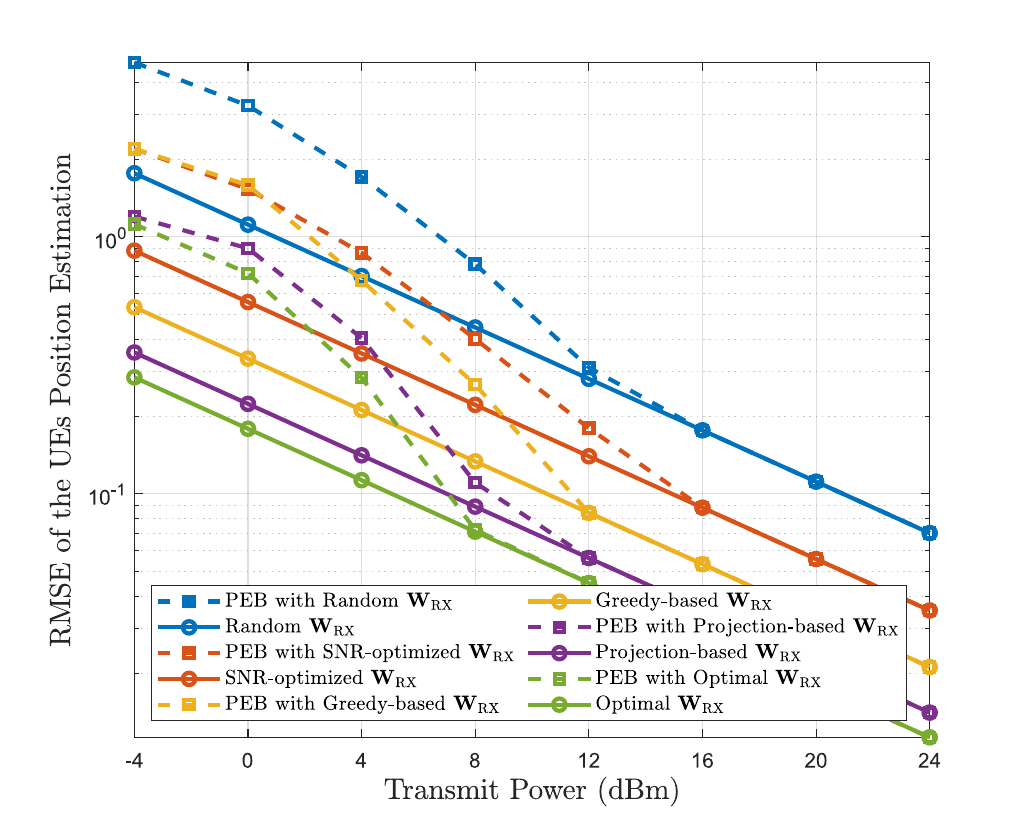}\vspace{-0.2cm}
\caption{\small{Mean RMSE of the {positions of the $U=3$ UEs} versus the transmit power $P_{\max}$ in dBm, considering an RX DMA with $N_{\rm RF} = 4$} microstrips each consisting of $N_{\rm E}=256$ phase-tunable metamaterials.}
\vspace{-0.4cm}
\label{fig:PEB}
\end{figure}
%width=0.92\columnwidth

\begin{figure}[!t]
\centering
\includegraphics[width=0.8\columnwidth]{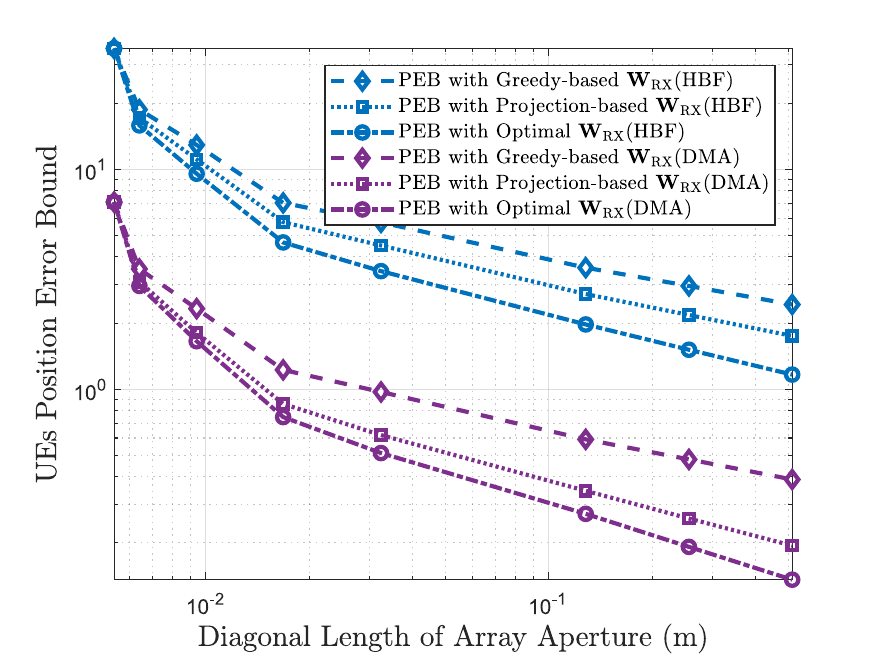}\vspace{-0.2cm}
\caption{\small{PEB versus the diagonal length in meters of the RX antenna panel, considering a DMA and an HBF with $\lambda/5$ and $\lambda/2$ element spacing, respectively, each with $N_{\rm RF}=4$, and $P_{\rm {\max}} = -4$ dBm.}}
\vspace{-0.4cm}
\label{fig:Ape}
\end{figure}

\begin{figure*}[t!]
    \centering
    \begin{subfigure}[t]{0.45\textwidth}
        \centering
        \includegraphics[scale=0.48]{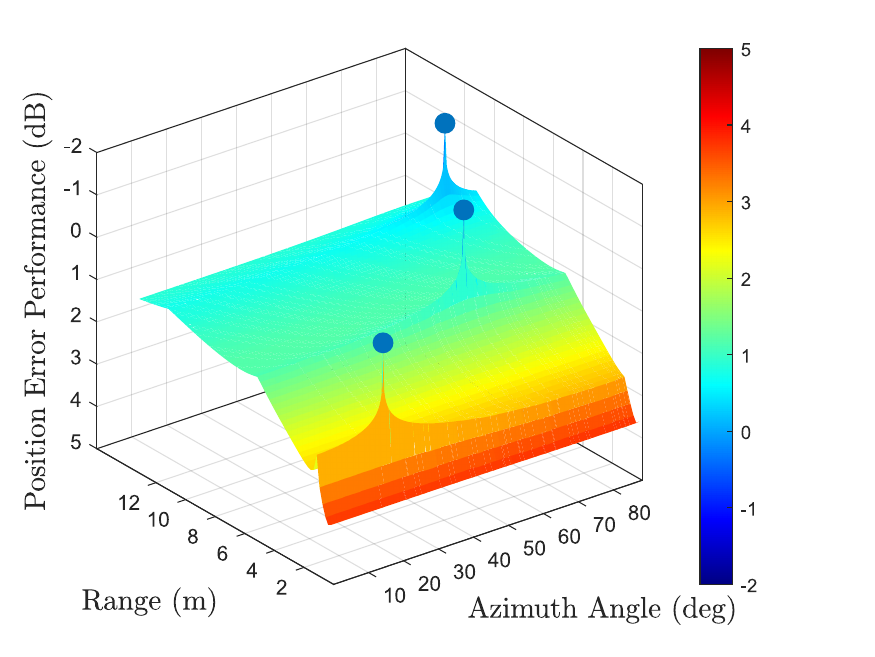}
        \caption{Projection-based solution.}
        \label{fig: proj}
    \end{subfigure}%
    \begin{subfigure}[t]{0.45\textwidth}
        \centering
        \includegraphics[scale=0.48]{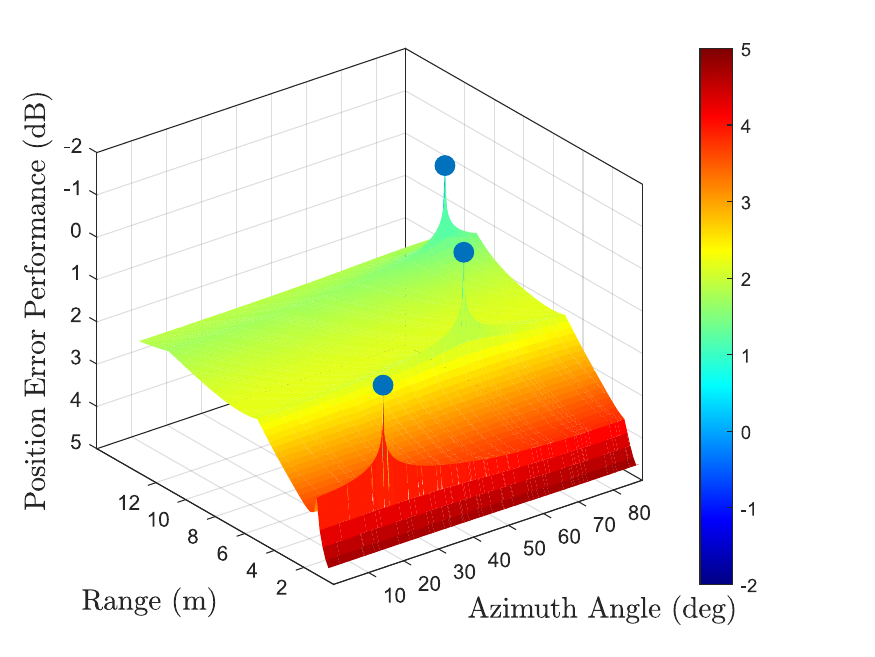}
        \caption{Greedy-based solution.}
        \label{fig: greed}
    \end{subfigure}
    \caption{\small{{Area-wide position estimation error with MLE for both proposed $\W_{\rm RX}$ designs (the vertical axes include error values in ascending order from the top to bottom), considering $U=3$ UEs each using $P_{\rm {\max}} = 20$~dBm for pilot transmission, and the DMA configuration of Fig.~\ref{fig:PEB}. The blue markers in both subfigures indicate the true positions of the UEs.}}}  %
    \label{fig:MLE}
\end{figure*}
In Fig.~\ref{fig:PEB}, the Root Mean Square Error (RMSE) of the UE's position and the PEBs of the proposed projection- and greedy-based $\W_{\rm RX}$ designs are compared with that of the localization-based DMA design in~\cite{FD_HMIMO_2023} maximizing the received Signal-to-Noise Ratio (SNR), a scheme with a random $\W_{\rm RX}$, and the optimal solution obtained through exhaustive search, when all considered to perform MLE at an RX DMA equipped with $N_{\rm RF} = 4$ microstrips each hosting $N_{\rm E}=256$ phase-tunable elements. As expected, all RMSE curves improve with increasing $P_{\rm {\max}}$ (i.e., SNR), converging to the respective PEBs. It is also shown that our optimization framework outperforms both that of~\cite{FD_HMIMO_2023} and randomized DMA reception, verifying the effectiveness of our near-field localization design. {Notably, the proposed projection- and greedy-based approaches exhibit comparable performance with the optimal RX BF solution, making our greedy-based one more preferable for low complexity implementations.}

Figure~\ref{fig:Ape} illustrates the PEB performance of both proposed designs together with that of the optimal solution for $\mathcal{OP}$ versus the diagonal length of the RX antenna
panel for two antenna architectures each with $N_{\rm RF}=4$ reception RF chains: DMA and partially connected Hybrid analog and digital BF (HBF) with phase shifters (via Section~\ref{Sec: 3} ignoring $\P_{\rm RX}$ and replacing $\mathcal{W}$ with the columns of the discrete Fourier transform matrix). The inter-element spacing within each microstrip of the former was chosen as $\lambda/5$, whereas, for the latter, $\lambda/2$ was the inter-element spacing of each single-RF-fed linear antenna array constituting the HBF. As expected, the denser element placement facilitated by the DMA architecture results in superior performance. It also shows that, as the size of the RX antenna panel increases, the gap between our projection-based design and the optimal solution (i.e., RX BF via exhaustive search) increases, with this gap being more pronounced for HBF for the same panel size.

{The MLE error performance over a grid of pairs of range and azimuth angle values is illustrated in Fig.~\ref{fig:MLE}, considering the RX DMA of Fig.~\ref{fig:PEB} and $P_{\rm max}=20$~dBm for each of the $U=3$ UEs. Estimations of the true positions of the UEs with random analog RX BF initializations were used for both our optimized $\W_{\rm RX}$ design approaches, which were then combined with MLE for the entire grid. It can be observed in the figure that the estimation error remains consistently low across short ranges and small azimuth angles, while it slightly increases at wider angles and larger ranges, where the SNR of the received pilots drops and the angular resolution decreases. It can be also seen that the isolated peaks with the smallest MLE values almost coincide with the blue markers signifying the true locations of the UEs, showcasing the effectiveness of both $\W_{\rm RX}$ designs. As demonstrated, on average, the projection-based solution outperforms the greedy-based one.}

\vspace{-0.1cm}
\section{Conclusion}
In this letter, we studied DMA-based reception for near-field localization, focusing on the THz frequency band. A novel expression for the CRB for the estimation of the UE range and angular parameters as well as the PEB were derived, and the former was deployed as the minimization objective for the design of the DMA analog RX BF weights. We capitalized on the partially-connected analog BF architecture of DMAs to reformulate its design objective into a Rayleigh quotient optimization problem with discrete constraints onto the analog BF weights, which was efficiently solved. Our simulation results verified the validity of our near-field localization analysis, showcasing the superiority of our presented localization-optimized DMA design over conventional ones, when all implemented together with MLE.

%The recent trend for extremely massive antenna arrays and high frequencies contributes to localization and sensing with increased angular and range resolution. In this letter, we focus on the emerging technology of Dynamic Metasurface Antennas (DMAs) and present a design framework for their analog beamforming weights targeting high accuracy near-field localization at the THz frequency band. We derive the Cram\'{e}r-Rao bound for user position estimation with a DMA-based receiver, which is then utilized as the optimization objective for the receiver's discrete tunable phases of its metamaterials. Leveraging the DMA structure, we reformulate its localization objective into a constrained Rayleigh quotient maximization problem, which is efficiently solved via a projection and a greedy-based schemes. Our simulation results verify the validity of our near-field localization analysis, showcasing the effectiveness of the proposed near-field localization designs over the optimum exhaustive search solution and state-of-the-art schemes.  

\vspace{-0.1cm}
\bibliographystyle{IEEEtran}
\bibliography{ms}
\end{document}